# Strain-Rate-Dependent Deformation Behavior of $Ti_{29}Zr_{24}Nb_{23}Hf_{24}$ High Entropy Alloys at Elevated and Room Temperature

Tangqing Cao [a, #], Wenqi Guo [b, #], Wang Lu [a], Yunfei Xue [a,*], Wenjun Lu [b], Jing Su [b], Christian H. Liebscher [b,*], Chang, Liu [b], Gerhard Dehm [b]

*a. School of Materials Science and Engineering, Beijing Institute of Technology, Beijing 100081, China*

*b. Max-Planck-Institut für Eisenforschung GmbH, Max-Planck-Straße 1, 40237 Düsseldorf, Germany*

*Corresponding author:

E-mail addresses:   xueyunfei@bit.edu.cn    liebscher@mpie.de

These authors contributed equally: Tangqing Cao, Wenqi Guo.

Key words: High-entropy alloy; strain rate; high temperature; deformation mechanism; kink bands.

**Abstract:**

We investigate the strain-rate-dependent mechanical behavior and deformation mechanisms of a refractory high entropy alloy, $Ti_{29}Zr_{24}Nb_{23}Hf_{24}$ (at.%), with a single-phase body-centered cubic (BCC) structure. High-temperature compression tests were conducted at temperatures from 700 to 1100°C at strain rates ranging from $10^{-3}$ to $10$ $s^{-1}$. A sudden stress drop after yield point was observed at higher temperatures and lower strain rates with the Zener-Holloman parameter, $lnZ$, in the range of 17.2-20.7. Such a softening behavior can be related to the interaction of dislocations with short-range clustering. However, at higher strain rates or lower temperatures ($lnZ>25.0$), kink bands were activated being responsible for the continuous strengthening of the alloy in competition with the softening mechanism. Systematic TEM investigations reveal that dislocation walls formed along {110} planes and dislocation loops were observed at a low strain of 6% at a high strain rate of 1 $s^{-1}$ and 800°C. Kink band induced dynamic recrystallization is evident upon further straining. On the other hand, at low strain rate of $10^{-3}$ $s^{-1}$ and 800°C, discontinuous recrystallization mechanisms become dominant with arrays of dislocations forming in front of the bulged boundaries of parent grains. These sub-grain boundaries eventually turn into high-angle grain boundaries. We also investigate the deformation mechanism of the alloy under extremely high strain rate ($10^3$ $s^{-1}$) at room temperature. The specimen exhibits extensive kink bands with arrays of dislocation walls. As further strained, multiple slip systems can be activated and the interaction of dislocation walls plays a vital role in the strain hardening of the alloy.

# 1. Introduction

High entropy alloys (HEAs) with multi-principal elements in relatively high concentrations (5–35 at.%) have expanded the space to design novel alloys with unprecedented properties [1-3]. Within this concept, a broad range of HEAs with a single face-centered cubic (FCC) or body-centered cubic (BCC) phase solid solution have been explored [4-6]. Refractory HEAs (RHEAs) containing high-melting refractory elements, such as Ta, Hf, Mo, or W, typically have a BCC structure[7]. They have attracted considerable attention due to their outstanding combination of room and high temperature strength[8]. The RHEAs showed yield stresses of >500 MPa at temperatures up to 600 ºC. The alloy families NbMoTaW and VNbMoTaW showed extraordinary high yield strength of >400 MPa up to temperatures of 1600 ºC even exceeding conventional Ni-base superalloys[7]. Thus, RHEAs are considered as a promising alloy system for high temperature applications. However, the dislocation mechanisms, deformation softening behavior and microstructure evolution in BCC-RHEA under high strain rate deformation especially at high temperatures remain elusive. The activation energy of dislocation motion in BCC-RHEAs is higher than the self-diffusion of corresponding high temperature BCC metals[9], which will affect the dominating deformation mechanisms. Thus, it is of a great interest to investigate the underlying temperature and strain rate dependent microstructure and dislocation evolution of RHEAs.

Generally it is observed that the yield strength of RHEA decreases with increasing temperature under quasi-static conditions[7]. The spread of the yield stresses reported in the literature strongly depends on alloy composition and processing history[10]. A comprehensive picture of the deformation mechanisms of BCC HEAs only exists for room temperature tests, where it is observed that ½<111> screw dislocations dominate plastic flow. The majority of dislocations remained straight suggesting a reduced mobility, but in some instance dislocation debris were found indicating that these dislocations were pinned by effects stemming from the supersaturated solid solution[11]. In the early stages of deformation, dislocations are arranged heterogeneously in deformation bands suggesting that high Peierls stresses are acting on their non-planar core. The deformation microstructure typically consists of a high density of bands with high dislocation density and dislocation free regions[10]. The high activation volume obtained at room temperature in a HfNbTaTiZr alloy does not fall in line with a pure Peierls mechanism, which is dominated by the nucleation of kink pairs. It is proposed that the associated enhancement in strength is related to local compositional fluctuations introducing modulations of the dislocation core structure. This leads to an increase of the nucleation barrier for double-kinks, thus increasing the barrier for dislocation slip[12]. The activation of double-kinks, strongly depends on temperature[13] and thus this thermally activated regime controls the deformation of RHEAs at elevated temperatures. On the other hand, dynamic recovery and/or recrystallization are considered as the main softening mechanism upon high temperature deformation at low strain rates [9, 14, 15]. Thus, the interplay of dislocation structure with the dynamic restoration of RHEAs deformed at elevated temperature and a broad range of strain rates needs to be further studied.

The yield stress of BCC RHEA solid solution alloys was also found to be sensitive to strain rate upon deformation at both room and elevated temperatures. At room temperature, a quasi-linear work hardening was observed in HfNbTaTiZr alloys for low strain rates, while strong softening occurs due to the formation of adiabatic shear bands at high strain rates[16]. Deformation of a quinary MoNbHfZrTi alloy revealed that at temperatures above 800 ºC, a strong increase of the yield stress was observed at low strain rates ($10^{-5}$ to $10^{-4} s^{-1}$), which tends to be more moderate at higher strain rates of $10^{-1} s^{-1}$[17]. Under extremely high strain rate deformation ($10^3$ $s^{-1}$), heterogeneously distributed deformation bands are observed in TiZrNbHfTa alloy at room temperature[16], and these deformation bands are considered as kink bands. At elevated temperatures, a sharp drop after the yield point was observed at lower strain rate ($10^{-4}$ $s^{-1}$ to $10^{-2}$ $s^{-1}$) in a TiZrNbHfTa alloy[9]. The sharp yield point drop was considered as dislocation de-pinning from solute atom atmospheres or short-range ordering or clustering. After yield stress drop, the flow stress showed a continuous softening, and necklace microstructures are formed near grain boundaries, and the deformation mechanisms are dominated by dynamic recrystallization (DRX). However, the deformation mechanism at both high strain rate and elevated temperature still remains elusive. Thus, revealing the evolution of the intrinsic deformation and dislocation structures and their effect on the mechanical properties will fill the gap in understanding the hot deformation behavior of BCC RHEA.

To unravel the underlying strengthening mechanism and anomalous yielding behavior of RHEA, we performed compression tests of a $Ti_{29}Zr_{24}Nb_{23}Hf_{24}$ alloy with a single phase BCC structure at elevated temperatures from 700 °C to 1100 °C and a broad range of strain rates from $10^{-3}$ $s^{-1}$ to 10 $s^{-1}$. We also performed compression tests at room temperature under extremely high strain rate of $10^3$ $s^{-1}$. The effect of strain rate on the yield behavior and the deformed microstructures such as dislocations and kink bands have been studied by electron backscatter diffraction (EBSD) and transmission electron microscopy (TEM). The strain-rate-dependent deformation mechanisms of the $Ti_{29}Zr_{24}Nb_{23}Hf_{24}$ alloy are revealed.

## 2. Methodology
### *2.1. Alloy compositions and thermomechanical processing design*

The nominal composition of the RHEA in the current study is $Ti_{29}Zr_{24}Nb_{23}Hf_{24}$ (at.%). The alloy was prepared by electromagnetic levitation melting a mixture of the constituent elements (purity＞99.9wt.%) in an argon atmosphere. The ingots were remelted at least three times to ensure chemical homogeneity of the alloy. The molten alloy was cooled and solidified in water-cooled copper crucible to obtain an alloy ingot with a cylindrical shape with a diameter of 100 mm and a height of 80 mm. Small pieces (25 × 16 × 80 $mm^3$) were extracted from the as-cast material and were homogenized at 1290°C for 2 h in argon protected atmosphere followed by water quenching. Cylindrical specimens with a diameter of 7 mm and a height of 10.5 mm were machined from the homogenized material. Hot compression tests for these samples were performed on a Gleeble-3500 thermo mechanical simulator. Five different forming temperatures (700, 800, 900, 1000, and 1100°C) and five different strain rates ($10^{-3}$, $10^{-2}$, $10^{-1}$, 1, and 10 $s^{-1}$) were used in hot compression tests. Images of the deformed samples tested at different strains are shown in Appendix 1. In order to minimize friction during hot deformation, a tantalum foil with a thickness of 0.1 mm was used between the sample and the dies. The deformation chamber was evacuated to 2.0 Pa before heating the sample. The chamber was then filled with inert argon, and hot-compression tests were carried out. Each specimen was heated to the desired temperature at a heating rate of 10 °C/s, and then kept for 240 s to eliminate the thermal gradient before loading. The sample was subsequently water quenched to room temperature in the Gleeble to directly maintain its deformed microstructure when the thermal compression test was completed. For room temperature testing the Gleeble-3500 was used at low strain rate ($10^{-3}$ $s^{-1}$) compression and a split Hopkinson bar was used for testing under high strain rate ($10^{3}$ $s^{-1}$) compression.

*2.2. Microstructure characterization*

The melting point of the as-cast alloy was measured by differential scanning calorimetry (DSC) in a NETZSCH STA 449 F3 instrument. Both heating and cooling rate are 10 °C/min, and 2 cycles were conducted for each specimen. The global crystal structure of the alloy was measured by X-ray diffraction (XRD) in a ISO-DEBYEFLEX 3003 diffractometer with a Co-Kα ($\lambda$= 1.79 Å) source operating at 40 kV and 40 mA between 20 and 130 ° (2 $\theta$) at a step size of 0.03° deg and a counting time of 30 s at each step. For microstructural analysis, the hot deformed specimens were sliced along the compression axis section. The cut surface was mechanically grinded with silicon carbide abrasive paper (P200 to P7000), and then polished by using 50 nm $SiO_2$ suspension to remove the deformation layer on the surface. Electron backscatter diffraction (EBSD) measurements were carried out using a JEOL 6490 scanning electron microscope (SEM). Electron channeling contrast imaging (ECCI) analysis was performed in a Zeiss-Merlin instrument[18]. Samples for transmission electron microscopy (TEM) observations were prepared in three steps, first by mechanical grinding down to a thickness of 100 μm, and second by mechanical dimpling down to 50 μm, and last by ion milling using Gatan PIPS 2 Model 695 with the energy ranging from 5 keV to 0.1 keV. TEM and scanning TEM (STEM) were conducted in a JEOL

## 3. Results and interpretation

### *3.1. Initial microstructure and composition*

The thermal stability and the initial microstructure of the homogenized alloy were characterized by XRD and EBSD. Fig. 1(a) exhibits the XRD result of the homogenized alloy. Only body-centered cubic (BCC) phase is detected in the homogenized $Ti_{29}Zr_{24}Nb_{23}Hf_{24}$ (at.%) alloy. The grain orientation and corresponding inverse pole figure (IPF) map is shown in Fig. 1(b). The grain structure consists of equiaxed grains with an average grain size of 361.2±143.0 μm (~40 grains are measured). Table 1 shows the wet chemical analysis result of the alloy. The measured chemical composition is consistent with the nominal composition of the designed alloy, which is nearly equimolar.

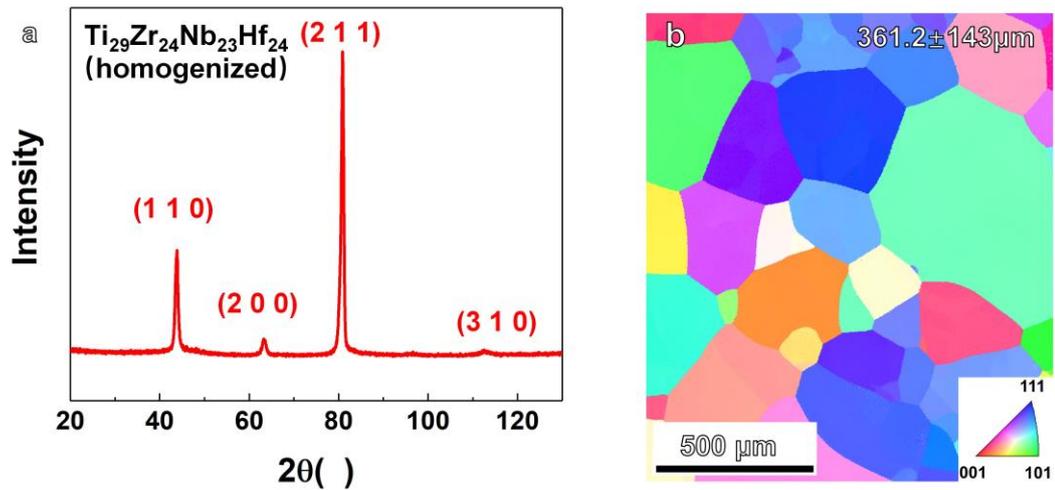

Figure 1. (a) XRD result of homogenized $Ti_{29}Zr_{24}Nb_{23}Hf_{24}$ alloy; (b) EBSD IPF map of homogenized $Ti_{29}Zr_{24}Nb_{23}Hf_{24}$ alloy.

Table 1. Wet chemical analysis of $Ti_{29}Zr_{24}Nb_{23}Hf_{24}$ alloy

| Sample  | Ti    | Zr    | Nb    | Hf    |
|---------|-------|-------|-------|-------|
| (at. %) | 28.87 | 23.84 | 22.98 | 24.31 |
| (wt. %) | 13.80 | 21.70 | 21.30 | 43.30 |

## 3.2. Mechanical properties

The temperature and strain rate dependent mechanical properties of the Ti$_{29}$Zr$_{24}$Nb$_{23}$Hf$_{24}$ (at.%) alloy have been investigated by uniaxial compression testing. Fig. 2(a) shows the true stress-strain curves of the alloys tested from 700 °C to 1100 °C at strain rates ranging from of $10^{-3}$ to 10 s$^{-1}$. A maximum strain of 0.2 is selected from the curve to make sure the true stress-strain data is valid. Actually the sample has good plasticity and maintains integrity with a strain up to 0.5, as shown in the Appendix 1. Stress-strain curves under all deformation conditions showed that the yield strength decreases as the temperature increases for a given strain rate. At a given temperature, the yield strength increases with increasing strain rate. This hardening behavior becomes more pronounced above 900 °C. Under certain conditions, for example at a strain rate of $10^{-3}$ s$^{-1}$ at 800 °C, the stress shows a sharp drop after the yield point reaching a plateau stress after a strain of ~10%, while a continuous hardening is observed after yielding at the strain rate of 1 s$^{-1}$ at 800 °C, indicating that the strain rate affects the deformation mechanism and flow stress.

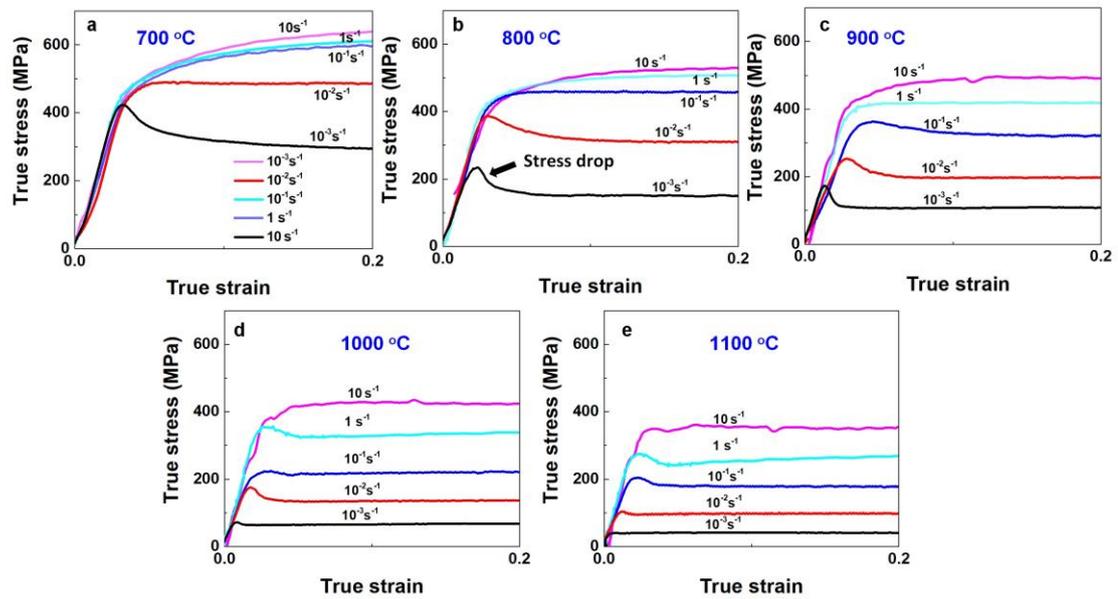

Figure 2 Compressive true stress-strain curves of homogenized Ti$_{29}$Zr$_{24}$Nb$_{23}$Hf$_{24}$ (at.%) alloy tested at strain rates between $10^{-3}$ s$^{-1}$ up to 10 s$^{-1}$ and different temperatures: (a) 700 °C; (b) 800 °C; (c) 900 °C; (d) 1000 °C; (e) 1100 °C.

To evaluate the global deformation behavior of the alloy and its strain rate and temperature dependence, the stress exponent $n$ and the activation energy $Q$ were obtained by fitting the Arrhenius-type constitutive equation according to [9]:

$$\dot{\varepsilon} = A\sigma^n \exp(-Q/RT) \tag{1}$$

Where $\dot{\varepsilon}$ is the strain rate (s$^{-1}$), $A$ a material constant, $\sigma$ the flow stress (MPa) at a certain strain (0.1), which is obtained from the stress-strain curves, $n$ the stress exponent, $Q$ the activation energy of dislocation motion for hot deformation (kJmol$^{-1}$), $R$ the gas constant (8.314 JK$^{-1}$mol$^{-1}$), $T$ the deformation temperature (K). Thus, the activation energy $Q$ can be estimated by the following equation [9],

$$Q = R[\partial \ln(\dot{\varepsilon})/\partial \ln(\sigma)]_T [\partial \ln(\sigma)/\partial(1/T)]_{\dot{\varepsilon}} \tag{2}$$

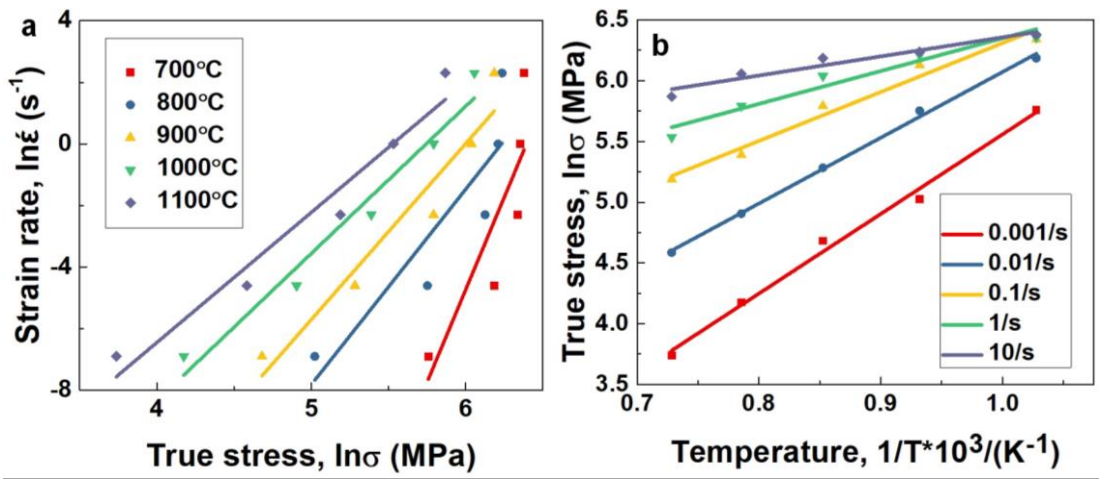

Figure 3. Flow stress analysis of the Ti$_{29}$Zr$_{24}$Nb$_{23}$Hf$_{24}$ (at.%) alloy: (a) ln ($\dot{\varepsilon}$) vs. (ln ($\sigma$)); (b) ln (ln ($\sigma$)) vs. 1/T.

The average activation energy (Q) was evaluated following equation 2 and the corresponding factors were extracted from the ln(strain rate) vs. ln(stress) and ln(stress) vs. 1/T [9] plots, as shown in Figs. 3(a) and (b), respectively. The stress exponent was obtained by determining the slope from the ln(strain rate) vs. (flow stress) graphs shown in Figs. 3(a), which ranges from 4.5±0.5 to 8.0±1.6. Figs.3 (b) shows the relationship between stress and temperature and the slope decreases from 1.5±0.23 to 6.5±0.33 as the strain rate increases. From the obtained T and rate dependent stress values we estimated the activation energy values $Q$ ranging from 56.1±4.2 kJ/mol to 432.3±13.3 kJ/mol for increasing strain rates from $10^{-3}$ s$^{-1}$ to 10 s$^{-1}$, respectively. The obtained activation energy is in a similar range to the self-diffusion of pure Nb (397 kJ/mol), but is about 2.2 to 3.4 times higher than that of pure Ti (198 kJ/mol), Hf (162 kJ/mol) and Zr (126 kJ/mol) [19]. It is interesting to note that the curves are not strictly linear. At lower strain rates (0.001 - 0.1/s), the relationship between ln(strain rate) and ln(stress) is nearly linear, which is comparable to that of a Ti$_{20}$Zr$_{20}$Nb$_{20}$Hf$_{20}$Ta$_{20}$ (at.%) alloy which exhibits an activation energy of ~258 kJ/mol[9]. This indicates that deformation in our alloy is governed by general dislocation climb during hot deformation at low strain rates. However, at high strain rates (0.1 - 10/s), the ln(strain rate) vs. ln(stress) relation departs from a linear behavior, suggesting that a change in deformation mode occurs. The Zener–Hollomon parameter (Z) is known to describe the combined effect of strain rate and temperature on flow stress of metals[20].

$$Z = \dot{\varepsilon} \cdot exp\frac{Q}{RT} \tag{3}$$

$$lnZ = \ln(A) + nln(\sigma) \tag{4}$$

When the $lnZ$ is in the range of 17.2~20.7, the yield drop is obvious, while as Z parameter is getting higher, there is a continuous strengthening after the yield point. For materials with high Peierls stress, the double-kink mechanism dominates dislocation motion[21]. At high temperature and low strain rate, dislocation climb makes an important contribution to the deformation mechanism due to thermal activation[21]. As the strain rate increases or temperature decreases (a higher $lnZ$), the thermal activation is insufficient, which could results in other type of deformation microstructures such as kink bands[22].

### 3.3. Deformed microstructure

Deformed microstructures of the alloy were characterized by EBSD and ECCI to study the effect of strain rate on the deformation features. The cross-section for microstructure observation is along the compression axis. Figs. 4 (a)-(d) show the IPF map of the deformed $Ti_{29}Zr_{24}Nb_{23}Hf_{24}$ alloy with a strain of 50% at 800 °C and 1100 °C with strain rates of $10^{-3}$ s$^{-1}$ and 1 s$^{-1}$. The deformed microstructure shows distinct features. At a temperature of 800 °C with the strain rate of $10^{-3}$ s$^{-1}$, a large amount of fine grains were observed along initial grain boundaries forming a conventional necklace-like structure [9, 17] after deformation. When the strain rate increases to 1 s$^{-1}$, the amount of fine grains is largely reduced and confined deformation takes place within the grains. The right part of Fig. 4(b) shows the {112} and {541} pole figures extracted from the black rectangular area. An evident strong misorientation around both {112} and {541} poles is observed, indicating the formation of kink bands[23]. At a temperature of 1100 °C, newly formed grains at the grain boundaries are more homogeneous and they have recrystallized to an average gain size of 85 μm (shown in the black rectangle of Figs. 4 (c)) deformed at a strain rate of $10^{-3}$ s$^{-1}$. When the strain rate is increased to 1 s$^{-1}$, necklace-like structures are again observed along initial grain boundaries, together with kink bands within the grains. The necklace-like structures have a softening effect while the kink bands leads to slight hardening. However, at $10^{-3}$ s$^{-1}$ there is almost no softening observed although some subgrains are formed. At the high temperature, yielding already sets in before the formation of DRX grains and then there is some dislocation hardening which counteracts the softening from the DRX grain formation. Strain rate has a great influence on the deformation mechanism, at low strain rate ($10^{-3}$ s$^{-1}$), dynamic recrystallization along grain boundaries is the dominant mechanism, while kink bands occur and participate in the deformation at high strain rate (1 s$^{-1}$) contributing to strain hardening.

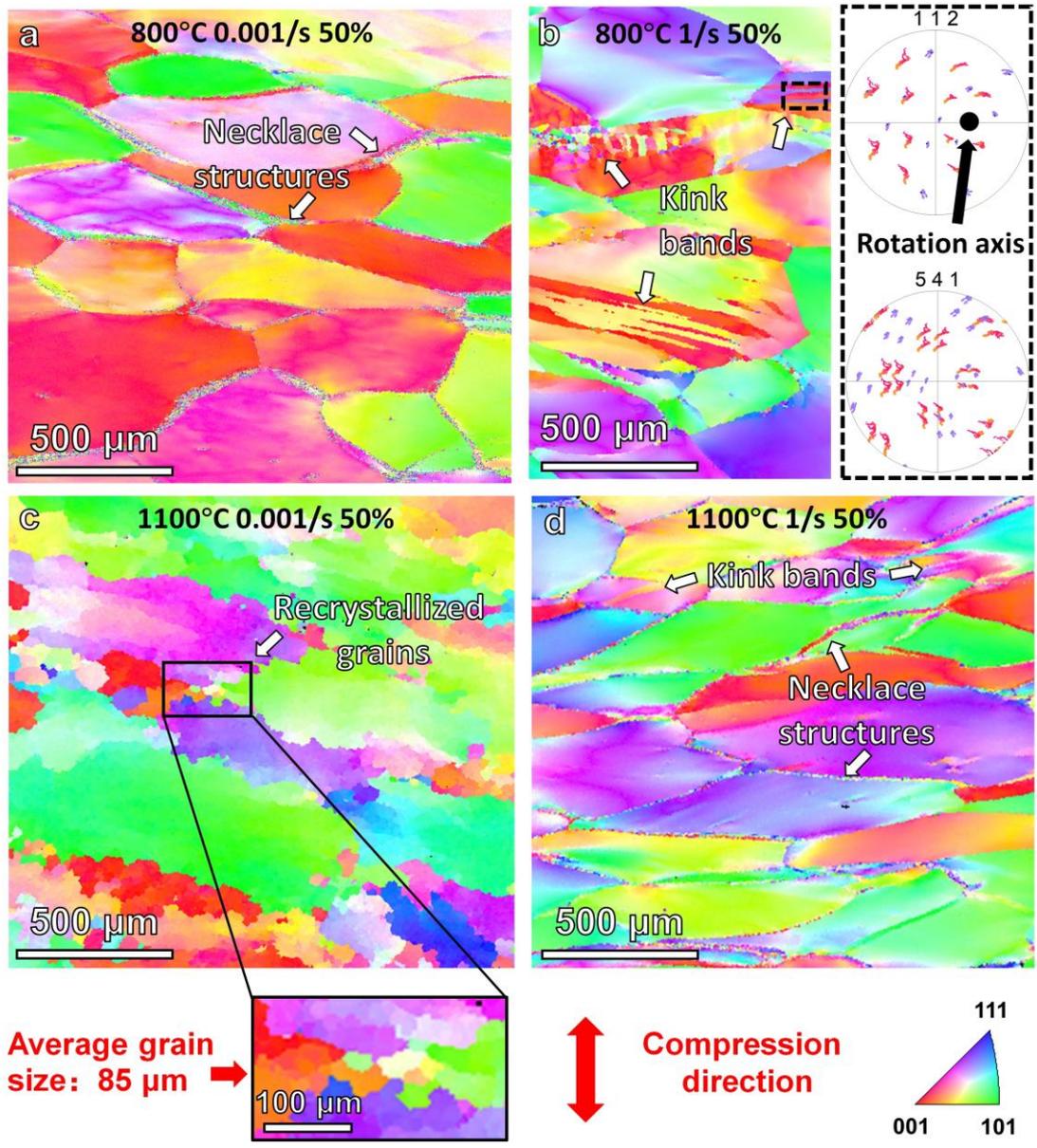

Figure 4. EBSD results of the deformed Ti$_{29}$Zr$_{24}$Nb$_{23}$Hf$_{24}$ alloy with ~50% strain at different temperatures and strain rates: (a) 800 °C and 10$^{-3}$ s$^{-1}$; (b): 800 °C and 1 s$^{-1}$; (c): 1100 °C and 10$^{-3}$ s$^{-1}$; (d): 1100 °C and 1 s$^{-1}$.

Figs. 5 (a) and (b) show the ECC image and EBSD IPF map of grain boundary of the alloy deformed to an engineering strain of 50% at 800 °C and at a low strain rate of 10$^{-3}$ s$^{-1}$. Refined subgrains/grains with average grain size of 8 μm are observed at the grain boundary after deformation, as shown in the white rectangle in Fig.5 (b). Most of these refined microstructures have a large misorientation with respect to the original bulk grains, which is shown in Figs.5 (c)~(e), and the point-to-origin misorientation angles between the refined grain (the rectangle in the center) and grain 1, grain 2 and grain 3 are 48°, 38° and 30°, indicating that the recovery and recrystallization starts from the grain boundary. The original grains are divided by subgrains/refined grains forming gradually from the grain boundary to the inner part to the grain.

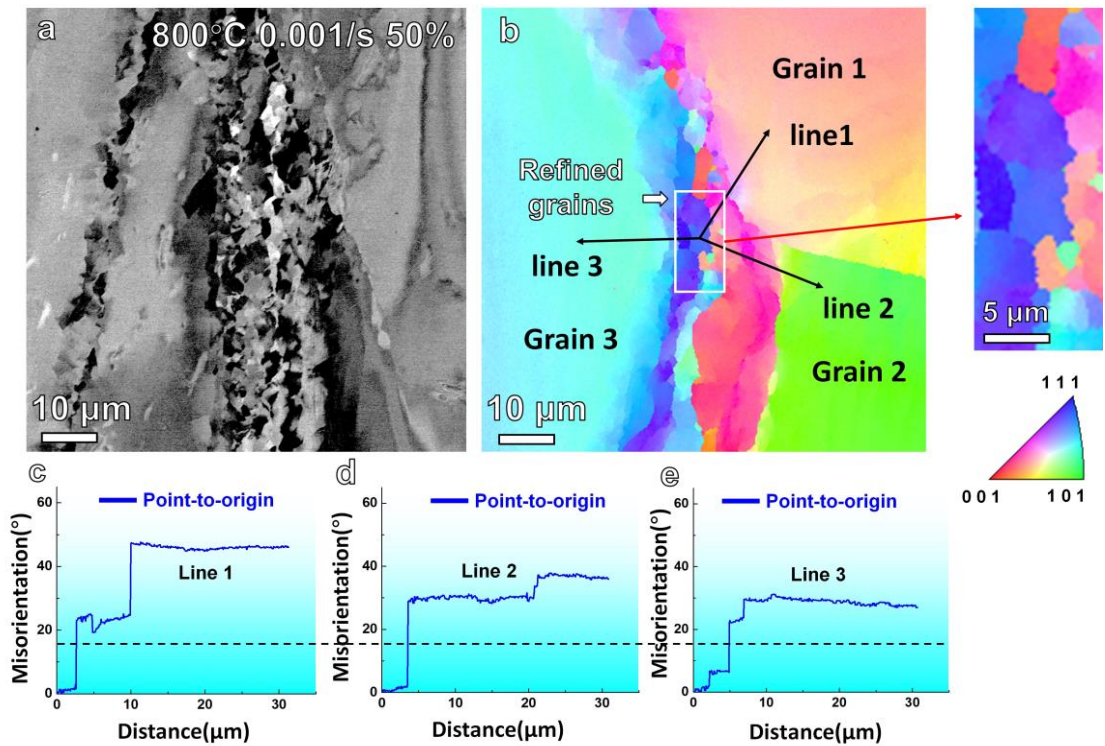

Figure 5. The deformed $Ti_{29}Zr_{24}Nb_{23}Hf_{24}$ alloy with ~50% strain at 800 °C and $10^{-3}$ $s^{-1}$: (a) ECC image; (b) EBSD IPF map(a); corresponding plots of the misorientation variation measured both with respect to the origin and from point to point of the lines in (b): (c) line 1; (d) line 2; (e) line 3.

Fig. 6 (a) shows the EBSD IPF map of the $Ti_{29}Zr_{24}Nb_{23}Hf_{24}$ (at.%) alloy deformed to an engineering strain of 50% at 800 °C and at a high strain rate of 1 $s^{-1}$. Large amounts of kink bands with (1 0 -1) habit plane are observed. Figs.6 (b)~(d) show the misorientation variation measured point-to-origin of the three lines in (a). The misorientation decreases with decreasing distance from the grain boundary. Is this due to the presence of the high angle grain boundaries, which generate strain gradient that limits the degree of misorientation. From Fig.4 to Fig.6 it can be clearly seen that strain rate has an obvious influence on the deformed microstructures of $Ti_{29}Zr_{24}Nb_{23}Hf_{24}$ (at.%) alloy. For a low strain rate of e.g. $10^{-3}$ s$^{-1}$, sub-grains with diameter of ~8 μm (shown in the white rectangle in Fig.5 (b) ) are observed along grain boundaries, indicating the initiation of recovery and grain refinement along high angle grain boundaries. At higher strain rate of 1 s$^{-1}$, kink bands with high misorientation gradient are observed within the grains.

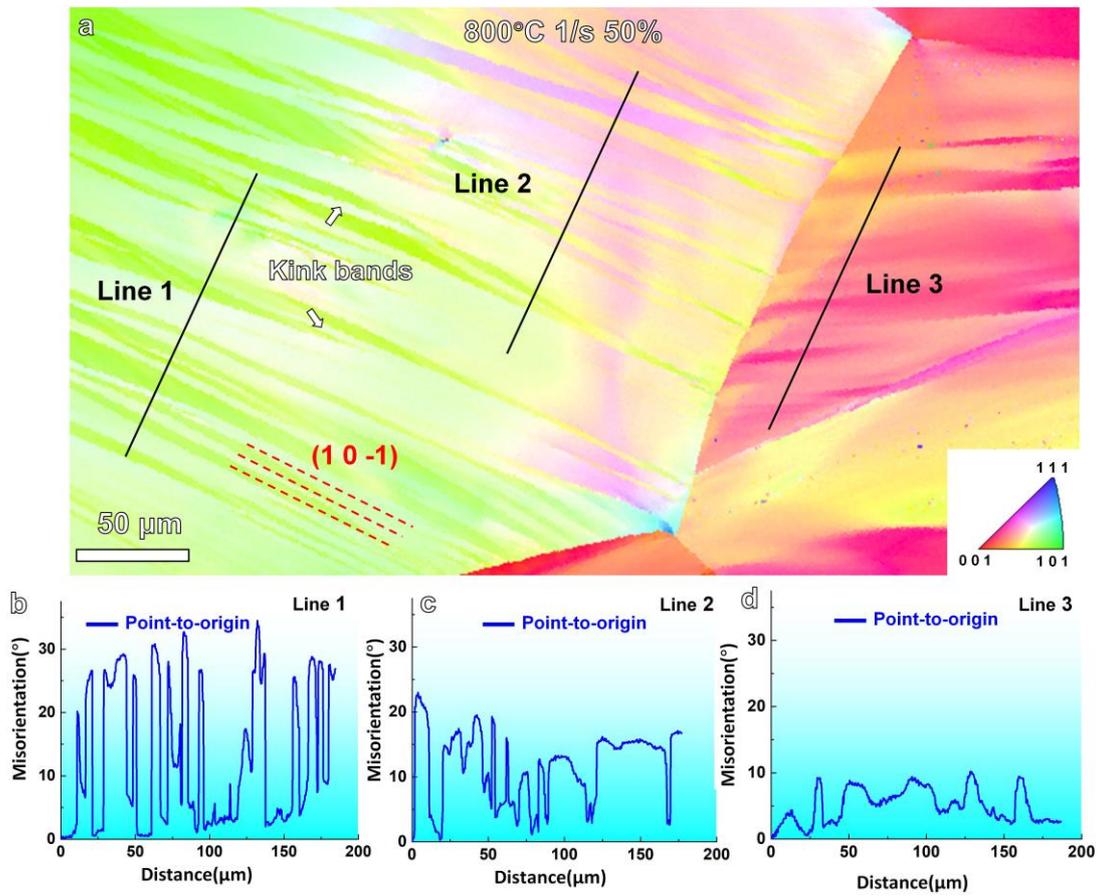

Figure 6. The deformed $Ti_{29}Zr_{24}Nb_{23}Hf_{24}$ alloy with ~50% strain at 800 °C and 1 s$^{-1}$: (a) ECC image; (b) EBSD IPF map; corresponding plots of the misorientation variation measured both with respect to the origin and from point to point of the lines in (b): (c) line 1; (d) line 2; (e) line 3.

### 3.4. Deformation mechanisms

To reveal detailed dislocation characteristics, STEM is used to analyze the samples after hot deformation at 800 °C with strain rates of $10^{-3}$ s$^{-1}$ and 1 s$^{-1}$. Figs. 7 (a) and (b) show bright field (BF) STEM images of the deformed alloy with ~50% strain at 800 °C and $10^{-3}$ s. The formation of sub-grain boundaries composed of a regular arrangement of dislocations are observed after hot deformation in close vicinity to the initial high angle grain boundaries, but dislocation arrays are also observed throughout the grain, which is consistent with the ECC and EBSD results shown in Fig.5. Under these conditions, deformation of the BCC high entropy alloy is dominated by heterogeneously distributed $\frac{a}{2}<111>$ dislocations (see Appendix 2) in the grain and dislocations arranging to form sub-grain boundaries[24, 25]. From these observations it can be speculated that the dislocations in the sub-grain boundaries nucleated at the high angle grain boundaries. In contrast, the sample deformed to a strain of ~50% at a strain rate of 1/s shows a dense and homogeneous distribution of dislocations with $\frac{a}{2}<111>$ Burgers vector (see Appendix 3), where strong dislocation-dislocation interactions are observed as shown in Figs. 7 (c) and (d). These act as barriers to further dislocation motion and promote work hardening at higher strain rates[26]. The above results can well explain the corresponding mechanical behavior. At 800 °C and $10^{-3}$ $s^{-1}$, dislocations are emitted from the GBs after yielding promoting the nucleation of subgrains at the GBs, which leads to softening and a sharp stress drop, as shown in Fig. 2. Upon further straining dislocation hardening from interacting dislocations within the grain counteracts the softening and a dynamic equilibrium state is reached leading to a plateau stress[20]. At higher strain rates (1 $s^{-1}$) the increased dislocation density and strong dislocation interactions result in continuous hardening after yielding instead of a stress drop and the formation of kink bands seems to play a vital role.

In order to remove the effect of temperature and isolate the intrinsic dislocation plasticity, as well as revealing slip traces and the local dislocation arrangement within the kink bands, EBSD, TEM and STEM are used to analyze the deformed structures in the $Ti_{29}Zr_{24}Nb_{23}Hf_{24}$ (at.%) alloy. Fig. 8 shows the EBSD results after deformation at strain rates of $10^3$ $s^{-1}$ and $10^{-3}$ $s^{-1}$ at room temperature. Under both strain rates, kink bands with {1 1 0} habit plane are observed at small strain (10%), and the maximum misorientation is ~5°, as shown in line 1 in Fig. 8(a) and line 4 in Fig. 8(d). As the strain increases, the fraction of kink bands largely increases and the maximum misorientation change increases to ~50° for strains above 30%. Compared to the test at extremely high strain rate ($10^3$ s$^{-1}$), the distribution of kink bands is more homogeneous at the same strain level compared to the test performed at lower strain rate ($10^{-3}$ s$^{-1}$).

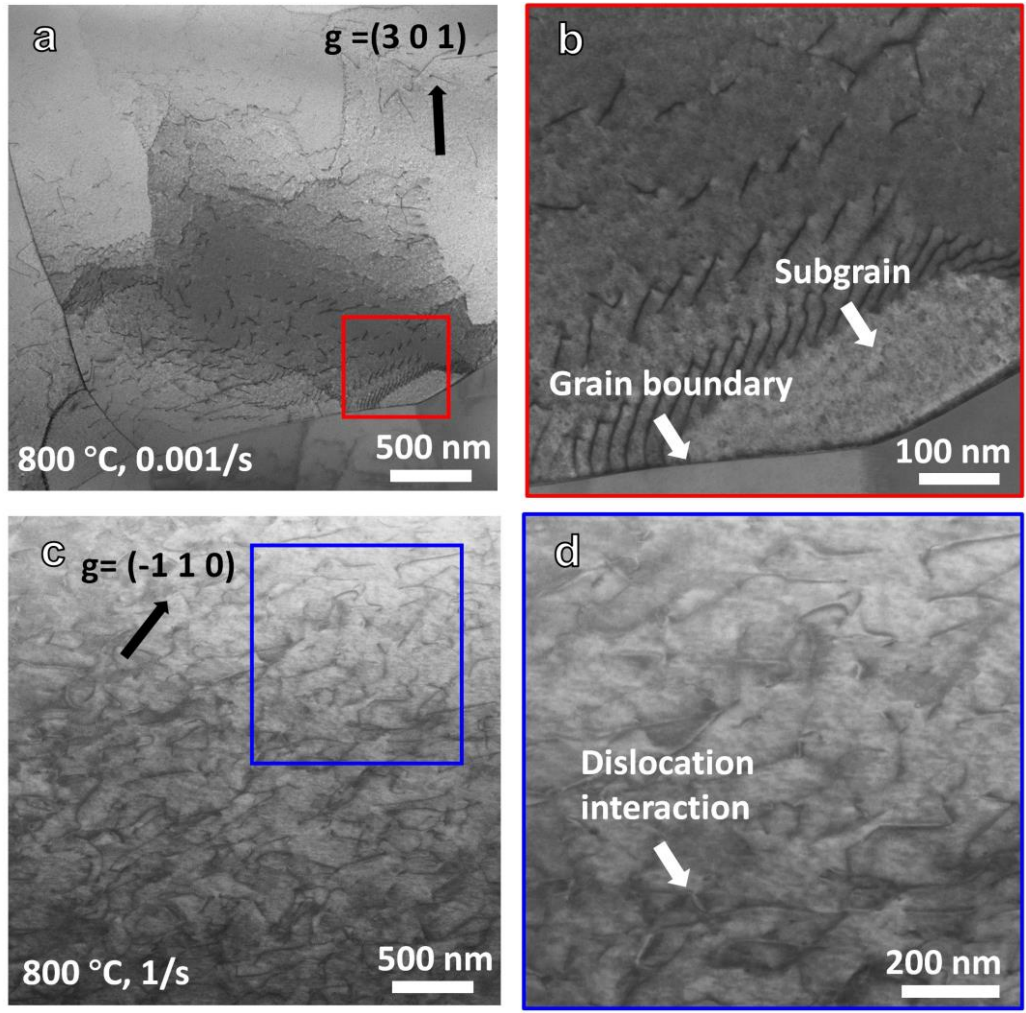

Figure 7. The STEM images of deformed Ti$_{29}$Zr$_{24}$Nb$_{23}$Hf$_{24}$ alloy with ~50% strain at 800 °C: (a) $10^{-3}$ s$^{-1}$; (b) enlarged image of red rectangular region in (a); (c) 1 s$^{-1}$; (d) enlarged image of blue rectangular region in (c).

Recent studies also reported the observation of kind bands in refractory high entropy alloys[16, 23], but the dislocation structure within these bands still remains elusive. Fig. 9 shows the dislocation structure of kink bands in the deformed Ti$_{29}$Zr$_{24}$Nb$_{23}$Hf$_{24}$ (at.%) alloy under different strain rates at room temperature. After a deformation strain of 5% at a strain rate of $10^3$ s$^{-1}$, localized dislocation walls on {1 1 0} planes are observed. As the strain increases to 10%, the dislocation structure remains heterogeneous with numerous dislocation walls still being present, but the activation of secondary slip and strong dislocation interactions are observed, which fits well with the EBSD result of Fig.8(a). Thus, the kink bands in $Ti_{29}Zr_{24}Nb_{23}Hf_{24}$ (at.%) alloy can be considered as an accumulation of dense dislocation networks aligned on {1 1 0} habit planes resulting from continuous slip. At low strain rate ($10^{-3}$ s$^{-1}$) and a strain of 10%, the dislocation distribution is more homogeneous and only weak signatures of deformation bands are detectable. Although strong dislocation interactions are observed, deformation at the lower strain rate is homogeneous, which is also in accordance with the EBSD results in Fig.8.

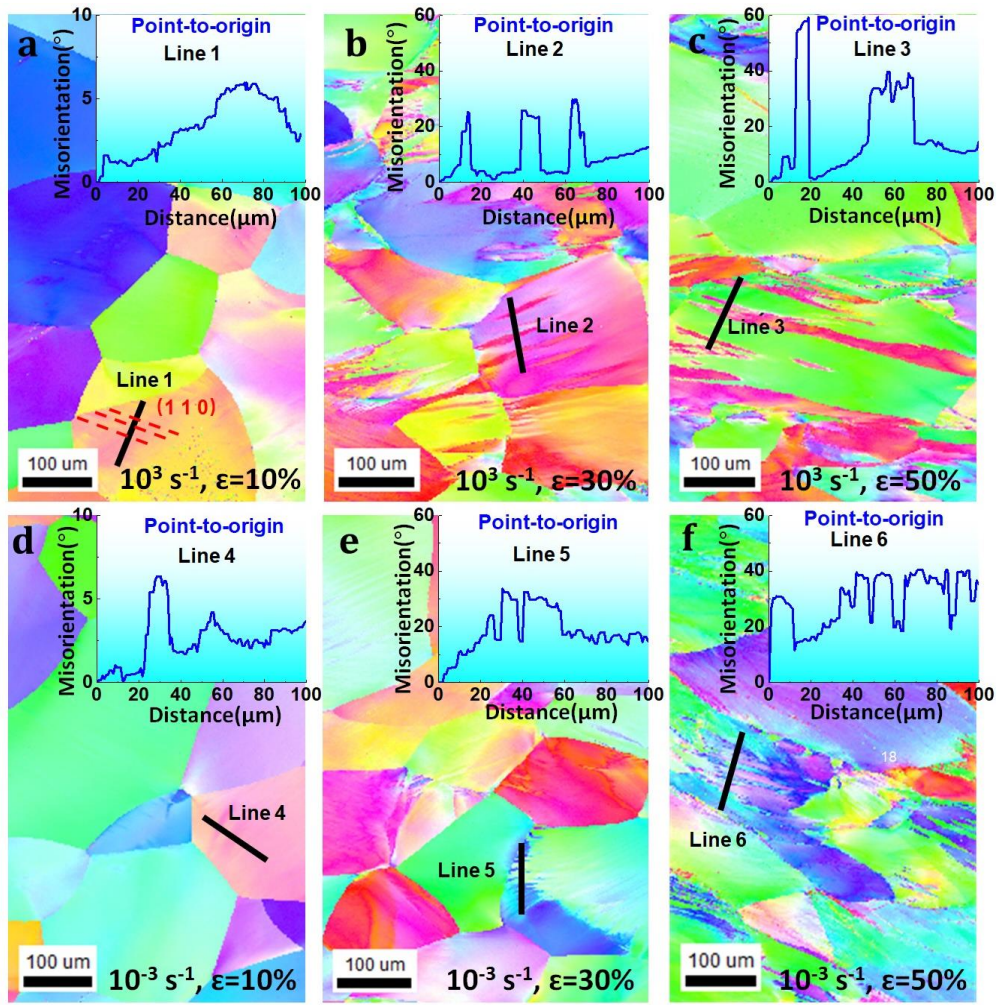

Figure 8. EBSD IPF maps of $Ti_{29}Zr_{24}Nb_{23}Hf_{24}$ refractory high entropy alloy deformed at room temperature with strain rate of $10^3$s$^{-1}$: (a) 10% strain; (b) 30% strain; (c) 50% strain; with strain rate of $10^{-3}$s$^{-1}$: (d) 10% strain; (e) 30% strain; (f) 50% strain.

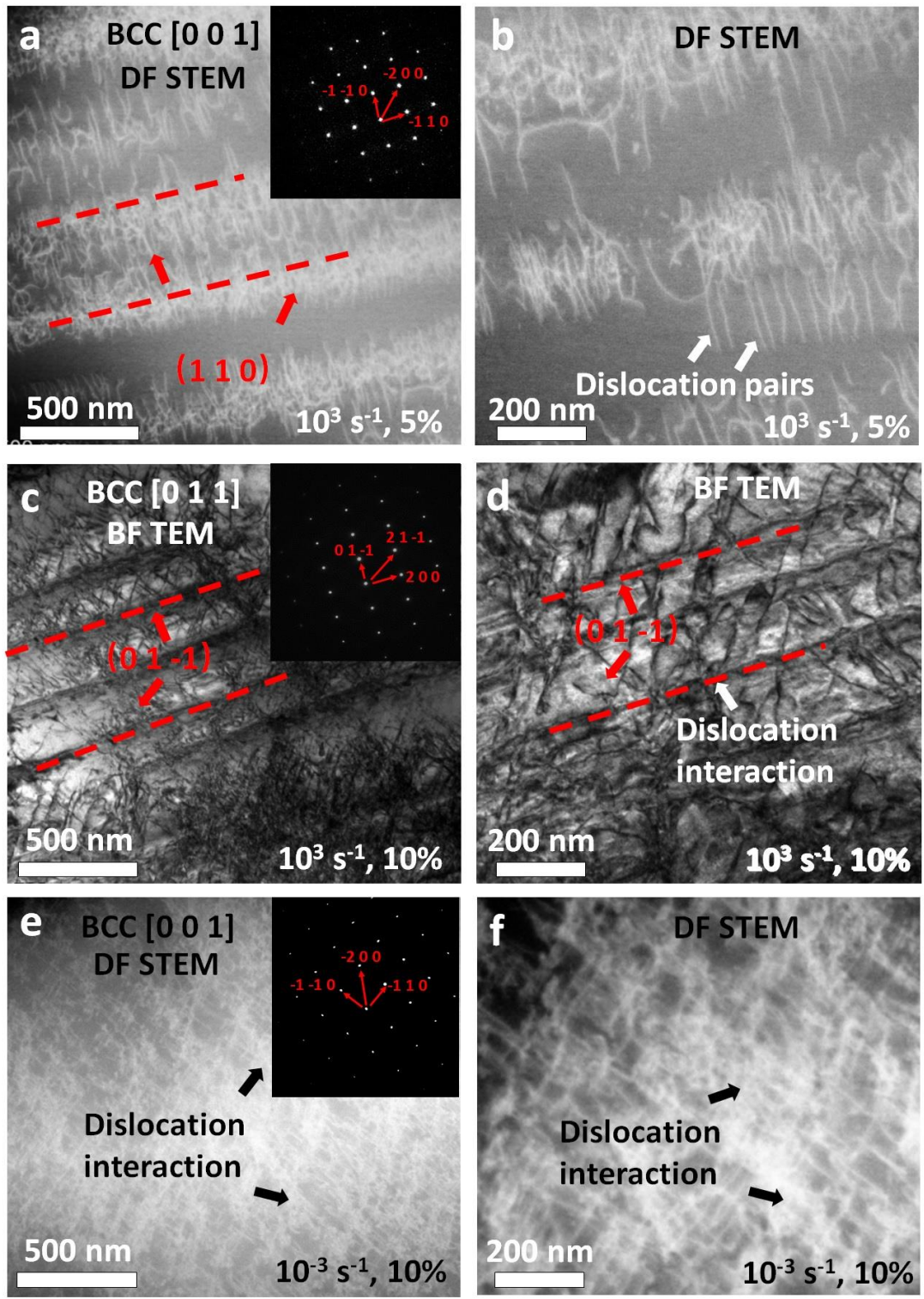

Figure 9. (a) and (b): Dark field STEM images and diffraction pattern of deformed $Ti_{29}Zr_{24}Nb_{23}Hf_{24}$ alloy with 5% strain and $10^3$ s$^{-1}$; (c) and (d); bright field TEM images and diffraction pattern with 10% strain and $10^3$ s$^{-1}$; (e) and (f); dark field STEM images and diffraction pattern with 10% strain and $10^{-3}$ s$^{-1}$.

## 4. Discussion

### *4.1. Effect of strain rate on yield strength*

During hot deformation, the plastic deformation of BCC alloys is controlled by dislocations motion[9] and dynamic restoration (recovery and recrystallization)[27, 28]. The resistance to dislocation motion can be separated into an athermal resistance, $\tau_a$, and a thermally-activated resistance, $\tau^*$. In the current study, we consider that $\tau^*$ is mostly due to the Peierls barrier for BCC metals[29, 30]. Based on this assumption, the total flow stress $\tau$ can be considered as the sum[31],

$$\tau = \tau^* + \tau_a \tag{5}$$

A general activation energy barrier model described in Eqn. (6) is often adopted to represent BCC alloys[31].

$$\Delta G = G_0[1 - (\frac{\tau^*}{\tau_0^*})^p]^q \tag{6}$$

where $\Delta G$ represents the free energy a dislocation must overcome by thermal activation, $\tau_0^*$ is the shear stress required to overcome the barrier at 0 K, $G_0$ is the free energy required to overcome the barrier when $\tau^*$ is zero, $p$ and $q$ are empirical parameters with $0 < p \leq 1$ and $1 \leq q \leq 2$. Strain rate and temperature are related though the Arrhenius expression of Eqn. (7)[32].

$$\dot{\varepsilon} = \dot{\varepsilon}_0 \exp(-\frac{\Delta G}{kT}) \tag{7}$$

where $k$ is the Bolzmann constant. Combining Eqn. (6) and Eqn. (7), the thermal flow stress can be obtained by Eqn. (8)[13].

$$\tau^* = \tau_0^*[1 - (\frac{kT}{G_0} \ln \frac{\dot{\varepsilon}_0}{\dot{\varepsilon}})^{1/q}]^{1/p} \tag{8}$$

Fig. 10(a)~(e) show the yield stress of $Ti_{29}Zr_{24}Nb_{23}Hf_{24}$ (at.%) alloy for strain rates between $10^{-3}$ s$^{-1}$ and 10 s$^{-1}$. The yield strength decreases quickly as temperature is increased, especially in the strain rate range ($10^{-3}$ s$^{-1}$ ~ $10^{-1}$ s$^{-1}$) and temperature range (700 °C ~900 °C). Under $10^{-3}$ s$^{-1}$, the yield strength of the $Ti_{29}Zr_{24}Nb_{23}Hf_{24}$ (at.%) alloy increases from 70 MPa to 420 MPa as the temperature decreases from 1100 °C to 700 °C, and the yield strength difference is 350 MPa, while for 10 s$^{-1}$, the yield strength difference is only 159 MPa within the same temperature range. At a low strain rate of $10^{-3}$ s$^{-1}$, the strength differences to pure Nb and Ta with BCC structure is ~80 MPa and ~90 MPa for 1100 °C and 700 °C, respectively [33]. At lower temperatures (700 ºC), the stress to move dislocations through a lattice with a high Peierls potential and possibly short range order (SRO) or clusters is higher than that in a pure metal. Moreover, SRO may as well affect the distribution of the Peierls potential surface and with this further impact double-kink formation and repulsion[34]. At 1100 ºC, the SRO domains might partially dissolve and dislocation climb processes become activated making it is easier for dislocations to bypass these local pinning points[9]. Thus, a contribution to the stress drop after yielding can be attributed to the unlocking of SRO domains [35-37].

For most strain rates, there is a plateau [13] of the yield stress at about 1100 °C, thus we use the stress at 1100 °C as the athermal stress and obtain the thermally activated stress through Eqn. (5), by fitting Eqn. (8) at different temperatures, which is shown in Fig. 10(f). From this, the Peierls stress at 0 K ($\tau_0^*$) can be deduced. In the current case $\tau_0^*$ is ~2200 MPa, which is about twice as high compared to pure Nb (1160 MPa) [13]. This implies that the compositional complexity of the supersaturated solid solution in the refractory high entropy alloy requires a higher stress for moving dislocations through the lattice.

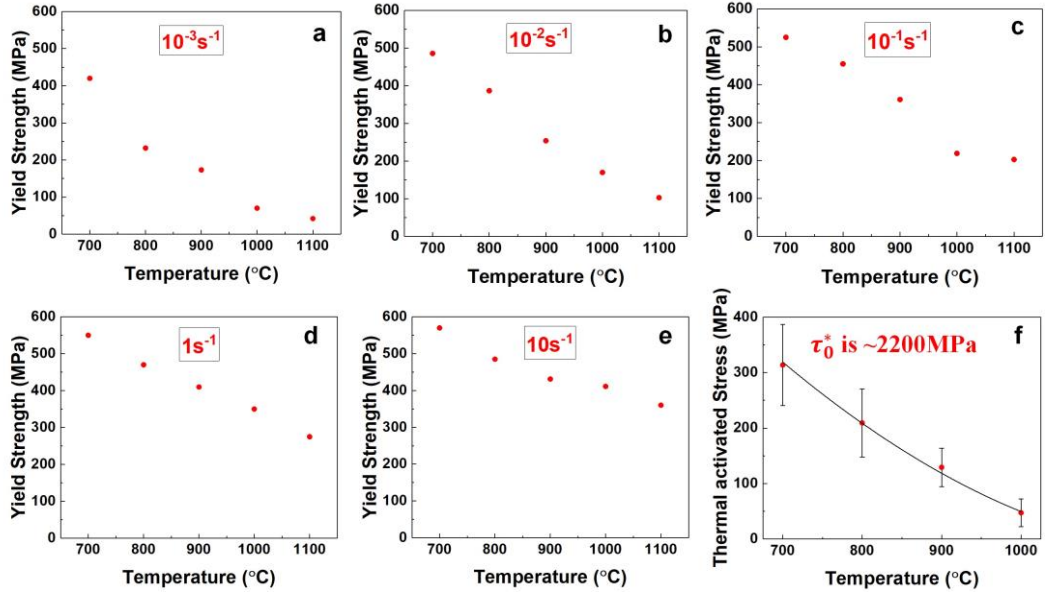

Figure 10. Yield strength of $Ti_{29}Zr_{24}Nb_{23}Hf_{24}$ alloy under the strain rate of: (a) $10^{-3}$ s$^{-1}$; (b) $10^{-2}$ s$^{-1}$; (c) $10^{-1}$ s$^{-1}$; (d) 1 s$^{-1}$; (e) 10 s$^{-1}$; (f) experimental and calculated thermally activated yield strength at different temperatures.

## *4.2. Deformation mechanism and kink bands*

For most BCC structured alloys and BCC high entropy alloys, deformation is controlled by dislocation motion, more particularly the motion of screw dislocations in BCC solid solutions is governed by a double-kink mechanism[38-40]. Hot compression is a thermally activated process[14, 16], in which both deformation temperature and strain rate will affect dislocation plasticity[41, 42]. At high temperature and low strain rate, dislocation climb may contribute to the deformation mechanism due to thermal activation[21], which is similar to that of FCC metals. As the strain rate increases or temperature decreases, the thermal activation is related to a regime where kink pairs are fully formed and the energetics is controlled by the repulsion of the kinks. At a strain rate of $10^{-3}$ s$^{-1}$ and 800 °C, when the stress reaches the critical stress for unlocking dislocations slip is suddenly activated contributing to softening of the material, which leads to a sharp stress drop. When these dislocations have arranged into sub grain boundaries dynamic recrystallization occurs and refined necklace-like grains are formed around the original grain boundaries. Upon further straining dislocation hardening counterbalances the softening and a dynamic equilibrium is reached leading to a plateau stress level. When temperature is increased to 1100 °C at $10^{-3}$ s$^{-1}$, dislocation nucleation and climb processes are promoted by thermal activation making it easier for dislocations to unlock or overcome SRO domains resulting in direct yielding and an almost negligible stress drop, as shown in Fig.2(e). At 1100°C and 1/s, a stress drop is observed, which is related to unlocking of dislocation from SRO due to the higher deformation rate and softening is promoted by the formation of necklace structure. But then, slight hardening is observed due to the formation of kink bands and dislocation walls which effectively reduce the mean free path for dislocation motion.

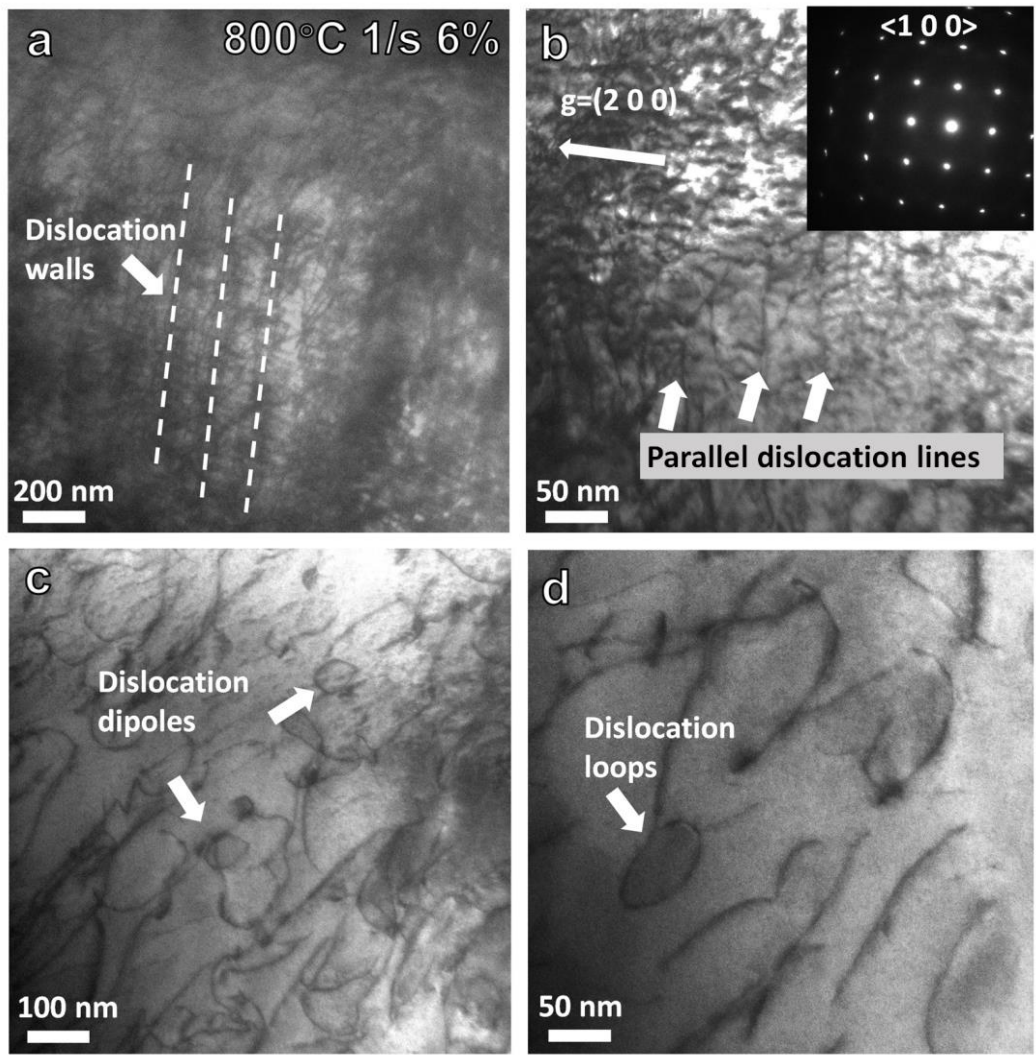

Figure 11. (a) TEM image and diffraction pattern of deformed Ti$_{29}$Zr$_{24}$Nb$_{23}$Hf$_{24}$ alloy at 800 °C with 6% strain and 1 s$^{-1}$; (b)-(d); bright field TEM images showing dislocation loops and dislocation dipoles.

At 800 ºC and a high strain rate of 1 s$^{-1}$, the dynamic recovery and dynamic recrystallization along grain boundaries are insufficient, while kink bands with arbitrary misorientation are observed after high temperature deformation leading to strain hardening, as shown in Figure 4(b). Figure 11(a) and (b) show the TEM images and diffraction patterns of a deformed Ti$_{29}$Zr$_{24}$Nb$_{23}$Hf$_{24}$ (at.%) alloy at 800 °C with 6% strain and 1 s$^{-1}$. Heterogeneously distributed dislocation walls were observed, which might contribute to the formation of kink bands. In Figures 11(c) and (d) the presence of dislocation loops indicates that dislocations are pinned by SRO. The kink mobility is disrupted by the SRO, which reduces the motion of lateral kinks and cross kinks due to the non-planar screw dislocation core[43], which finally contributes to a higher stress needed to move dislocations. In previous studies, the presence of debris in the microstructure was only observed at room temperature. However, in the current study, such debris is observed at elevated temperature due to the effect of high strain rate deformation. The increased strain rate reduces the effect of thermal activation, and SRO has a pinning effect on dislocations, finally resulting in the formation of heterogeneous kink bands. As the strain increases, EBSD observations show that the fraction of kink bands and local misorientation largely increases.

Kink bands are local deformation bands with an arbitrary degree of crystallographic rotation[22], which were commonly seen in Zirconium[44] and Magnesium[45]. During kink band formation, randomly distributed dislocations arrange into dislocation walls to reduce long range stress fields [46]. Our TEM observations revealed that at the beginning stage of high strain rate deformation of the Ti$_{29}$Zr$_{24}$Nb$_{23}$Hf$_{24}$ (at.%) alloy, dislocations partially arrange into regularly distributed dislocation walls with {1 1 0} habit plane. At medium strain levels, regularly distributed dislocation walls are observed to form on {1 1 0} planes, which could be related to the existence of SRO in the alloy, since it will promote the planarity of dislocation slip[47]. The activation of secondary slip at higher strains leads to strong dislocation interactions, which contributed to strain hardening. In comparison, randomly distributed dislocations are observed in pure Ta with BCC crystal structure[48] at the same deformation strain and strain rate. Thus, the increased planar slip of dislocations is a distinctive characteristics of high entropy alloys with BCC crystal structure. As the deformation continues, the dislocations remain heterogeneous and multi-plane slip and dislocation interactions are formed. Thus, the accumulation of dislocation networks finally results in the formation of macroscopic kink bands in the $Ti_{29}Zr_{24}Nb_{23}Hf_{24}$ (at.%) alloy, which play a vital role in the continuous strain hardening at high strain rates.

**5. Conclusions**

Strain-rate-dependent deformation behaviors of $Ti_{29}Zr_{24}Nb_{23}Hf_{24}$ high entropy alloy with single-phase BCC structure were investigated. Uniaxial compression was carried out at different strain rates ranging from $10^{-3}$ $s^{-1}$ to 10 $s^{-1}$ and temperatures ranging from 700 °C to 1100 °C, as well as compression at extremely high strain rate of $10^3$ $s^{-1}$ and room temperature. The effect of strain rate on the yield behavior, macroscopic deformation features and dislocation configurations were investigated.

1. Compared to pure metals, temperature has a more obvious effect on the yield strength of the $Ti_{29}Zr_{24}Nb_{23}Hf_{24}$ (at.%) high entropy alloy. Under $10^{-3}$ $s^{-1}$, the yield strength of $Ti_{29}Zr_{24}Nb_{23}Hf_{24}$ alloy increases from 70 MPa to 420 MPa as the temperature decreases from 1100 °C to 700 °C with a yield strength difference being 350 MPa, which is much larger than that of pure Nb and Ta (80-90 MPa). The compression tests performed at different strain rates and temperatures lead to an estimated Peierls stress of 2200 MPa at 0 K, which is about twice the Peierls stress at 0K for pure Nb, indicating that $Ti_{29}Zr_{24}Nb_{23}Hf_{24}$ alloy has a more obvious strain rate effect on stress than the pure metal. At 800 °C, the stress after yield shows a sharp drop at the strain rate of $10^{-3}$ $s^{-1}$, while a continuous hardening is observed after yielding at the strain rate of 1 $s^{-1}$.
2. The high temperature deformation mechanism of $Ti_{29}Zr_{24}Nb_{23}Hf_{24}$ (at.%) high entropy alloy is controlled by dislocation motion. At 800 °C and $10^{-3}$ $s^{-1}$, the yield drop is promoted by dislocation unlocking from SRO and the initiation of formation of a subgrain structure along grain boundaries. As deformation continues, dynamic recrystallization occurs at the grain boundaries and refined grains are formed, finally growing into recrystallized grains.

3. At 800 °C and higher strain rate condition (1 s$^{-1}$), macroscopic kink bands are first found to participate in the hot deformation process of BCC high entropy alloy. The dynamic recovery and dynamic recrystallization along grain boundaries is insufficient at high strain rates, and planar dislocation slip and the formation of dislocation walls occurs within the grain. As deformation continues, kink bands form and result in a continuous strain hardening.

4. Kink bands also play a dominant role at room temperature. At 10$^3$ s$^{-1}$, heterogeneously distributed dislocation walls on {1 1 0} planes are formed at the initial deformation stage of the Ti$_{29}$Zr$_{24}$Nb$_{23}$Hf$_{24}$ (at.%) high entropy alloy. With increasing strain, multi-plane slip promotes strong dislocation interactions which finally evolve into kink bands. The misorientation between kink bands increases with increased strain. At low strain rate of 10$^{-3}$ s$^{-1}$, a high density of mostly homogeneously distributed dislocations and dislocation networks are found.

**Acknowledgements**

Christian H. Liebscher and Gerhard Dehm acknowledge funding from German Science Foundation DFG for the SPP 2006 CCA-HEA. The authors would like to gratefully acknowledge the kind support of B. Breitbach, V. Kree, P. Watermeyer, S. Reckort, M. Nellessen, K. Angenendt and D. Kurz at the Max-Planck-lnstitut für Eisenforschung.

**Appendix**

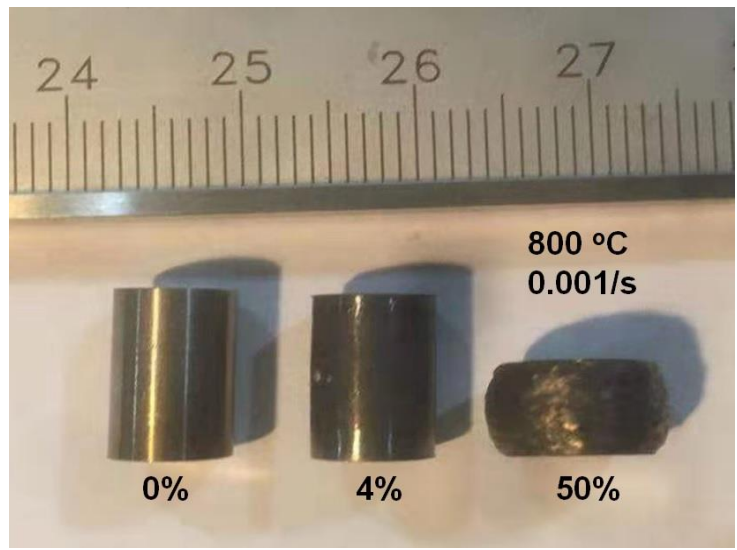

Appendix 1. Deformed $Ti_{29}Zr_{24}Nb_{23}Hf_{24}$ refractory high entropy alloy samples with different strains.

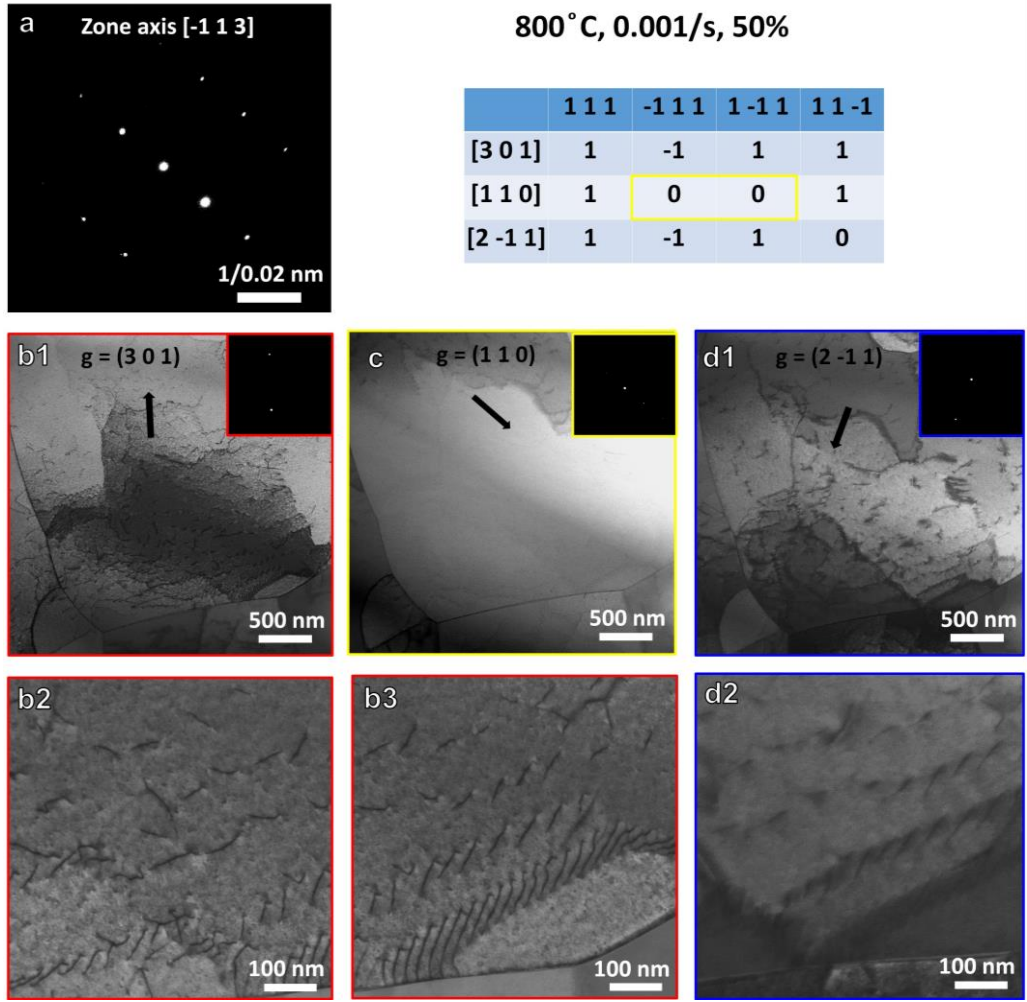

Appendix 2. The STEM images and diffraction patterns of deformed $Ti_{29}Zr_{24}Nb_{23}Hf_{24}$ alloy with ~50% strain at 800 °C and $10^{-3}$/s: (a) diffraction pattern of [-1 1 3] zone axis; dislocations (b1) and (b2) with g vector of [3 0 1]; (c1) and (c2) with g vector of [110]; (d1) and (d2) with g vector of [2 -1 1].

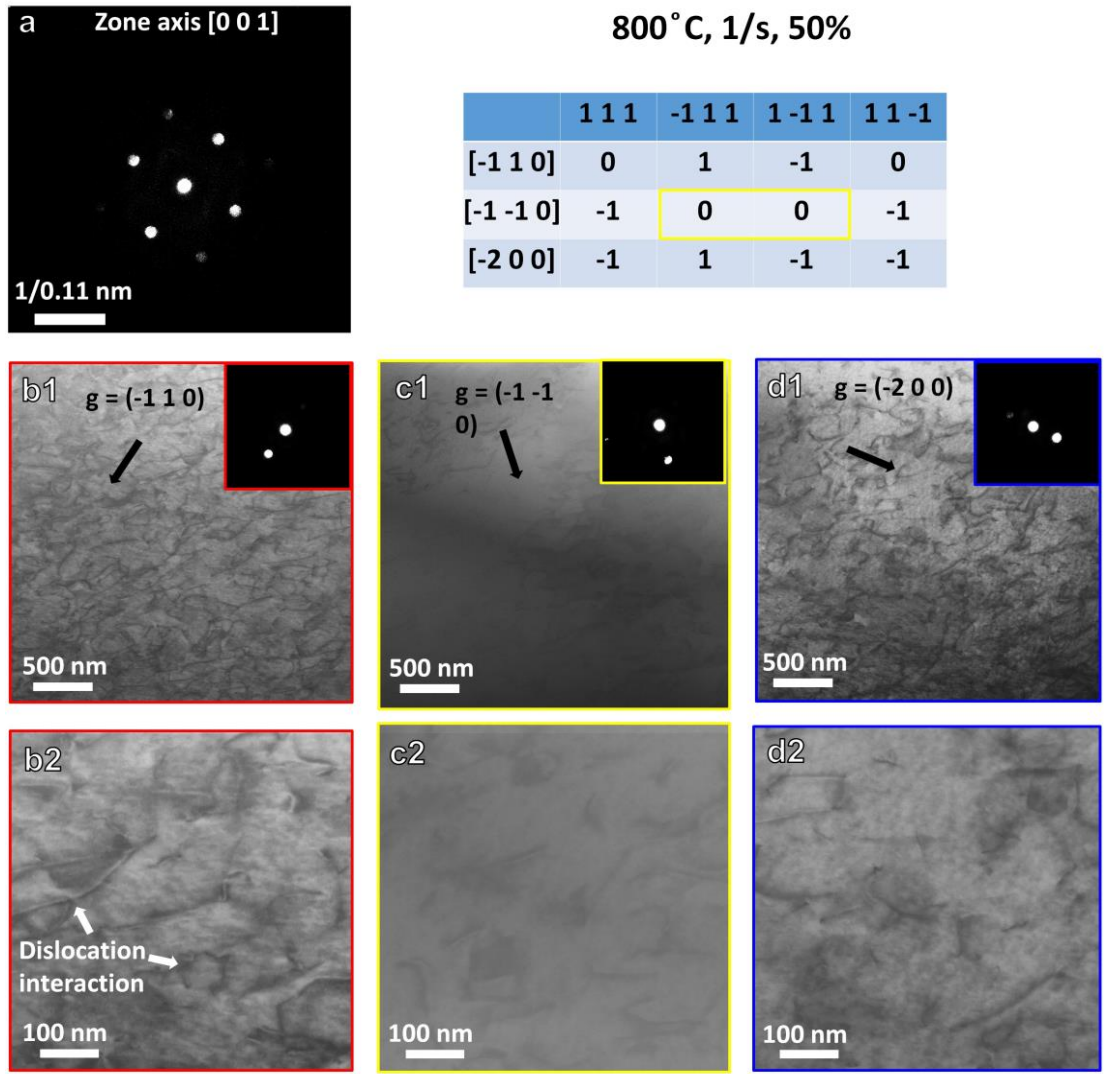

Appendix 3. The STEM images and diffraction patterns of deformed Ti$_{29}$Zr$_{24}$Nb$_{23}$Hf$_{24}$ alloy with ~50% strain at 800 °C and 1/s: (a) diffraction pattern of [0 0 1] zone axis; dislocations (b1) and (b2) with g vector of [-1 1 0]; (c1) and (c2) with g vector of [-1 -1 0]; (d1) and (d2) with g vector of [-2 0 0].